# Thermal Conductivity and Phonon Transport in Suspended Few-Layer Hexagonal Boron Nitride

*Insun Jo[1], Michael Thompson Pettes[2], Jaehyun Kim[2], Kenji Watanabe[3], Takashi Taniguchi[3], and Zhen Yao[1], Li Shi\*,[2]*

[1]Department of Physics and [2]Department of Mechanical Engineering, The University of Texas at Austin, Austin, TX 78712, USA

[3]National Institute for Materials Science (NIMS), Tsukuba, Ibaraki 305-0044, Japan

**ABSTRACT** The thermal conductivity of suspended few-layer hexagonal boron nitride (h-BN was measured using a micro-bridge device with built-in resistance thermometers. Based on the measured thermal resistance values of 11-12 atomic layer h-BN samples with suspended length ranging between 3 and 7.5 μm, the room-temperature thermal conductivity of a 11-layer sample was found to be about 360 Wm$^{-1}$K$^{-1}$, approaching the basal plane value reported for bulk h-BN. The presence of a polymer residue layer on the sample surface was found to decrease the thermal conductivity of a 5-layer h-BN sample to be about 250 Wm$^{-1}$K$^{-1}$ at 300 K. Thermal conductivities for both the 5 layer and the 11 layer samples are suppressed at low temperatures, suggesting increasing scattering of low frequency phonons in thin h-BN samples by polymer residue.

Both hexagonal boron nitride (h-BN) and graphite belong to the family of hexagonal layered materials. Although the lattice constants are similar for the two materials, the electronic properties are completely different.[1] While graphite is a semi-metal, breaking of the sub-lattice symmetry in h-BN results in a wide direct band gap of 5.8 eV, which enables its use as a dielectric layer in electronic devices as well as a lasing material in the ultra-violet spectrum.[2] Because of the clean and atomically smooth surface of h-BN, the electron mobility of graphene electronic devices supported on exfoliated few-layer h-BN has been found to be an order of magnitude higher than those supported on silicon dioxide ($SiO_2$).[3] In addition, the basal-plane thermal conductivity of a pyrolytic h-BN bulk sample has been reported to be as high as 390 $Wm^{-1}K^{-1}$ at room-temperature.[4] This value is 280 times higher than that for the $SiO_2$ dielectric used in current-generation silicon electronic devices, and is lower than only few dielectrics such as diamond. Although the thermal conductivity along the *c*-axis of h-BN is as low as 2 $Wm^{-1}K^{-1}$ because of the anisotropic layered crystal structure,[5] the high basal-plane thermal conductivity can be used to enhance lateral heat spreading when few-layer h-BN is used as the dielectric support for graphene electronic devices as well as other future-generation thin-film devices made of silicon, conducting polymers, or other novel layered semiconductors.[6] In contrast, although the basal plane thermal conductivity of graphite can be even higher than that reported for h-BN bulk crystals, the semi-metallic nature of graphite and few-layer graphene can result in electrical shorting when they are used as a heat spreader directly beneath the active layer in electronic devices. Despite the attractive thermal and electrical properties of h-BN for thermal management, recent studies of thermal transport in two-dimensional materials have been limited to single and few-layer graphene.[7, 8] Although a theoretical study of phonon transport in few-layer h-BN has been reported recently,[9] experimental data on the thermal transport properties in this system are lacking.

Here we report thermal transport measurements of suspended few-layer h-BN samples. Because the intensity of the Raman peaks in few-layer h-BN is very weak compared to those in few-layer graphene, micro-Raman thermometry methods developed for thermal measurements of graphene[7, 10, 11] are not applicable for h-BN. Hence, we have established a method to transfer and suspend few-layer h-BN samples onto micro-bridge devices with built-in resistance thermometers for thermal transport measurements. With this method, few-layer h-BN samples of different suspended lengths and layer thicknesses have been measured to obtain the thermal conductivity and the thermal contact resistance in the temperature range between 45 and 450 K.

The micro-bridge device is modified from an earlier design for thermal transport measurements of supported graphene.[12] The current design consists of a suspended structure fabricated from a 500-nm-thick $SiN_x$ film, on which four 10/70-nm-thick Cr/Pt metal lines are patterned. The few-layer h-BN sample was transferred and suspended onto the central rectangular frame made of $SiN_x$ beams according to the procedure shown in Fig. 1(a-d). In this procedure, few-layer h-BN was first exfoliated from h-BN powder crystals onto a silicon substrate covered with ~290 nm thick thermal oxide, which provides sufficient optical contrast for identifying few-layer h-BN (Fig. 1(a)). After a poly(methyl methacrylate) (PMMA) layer was spin coated on the sample and patterned with the use of electron-beam lithography (EBL), the exposed region of the few-layer h-BN flake was etched by $CF_4/O_2$ plasma to form a rectangular h-BN ribbon. Subsequently, EBL and metal lift-off process were used to deposit Au registration marks around the patterned h-BN [Fig. 1(b)]. A ~1.5 μm thick PMMA film was then spin-coated onto the substrate. After the sample was placed in a 1% hydrofluoric acid solution, the interface oxide was etched so that the PMMA layer was detached from the substrate. The few-layer h-BN sample and the gold alignment marks were attached to the PMMA film that floated on top of the solution. Upon thorough rinsing with deionized water, the wet PMMA film was manually aligned to the micro-bridge device under an optical microscope [Fig. 1 (c,f)]. After the deionized water was evaporated, the device was annealed at 150°C in vacuum to increase the adhesion between the few-layer h-BN and the micro-bridge device. The PMMA film was subsequently dissolved in acetone heated to about 60°C and then dried. Figure 1(g) shows a completed device consisting of a suspended 11-layer h-BN sample.

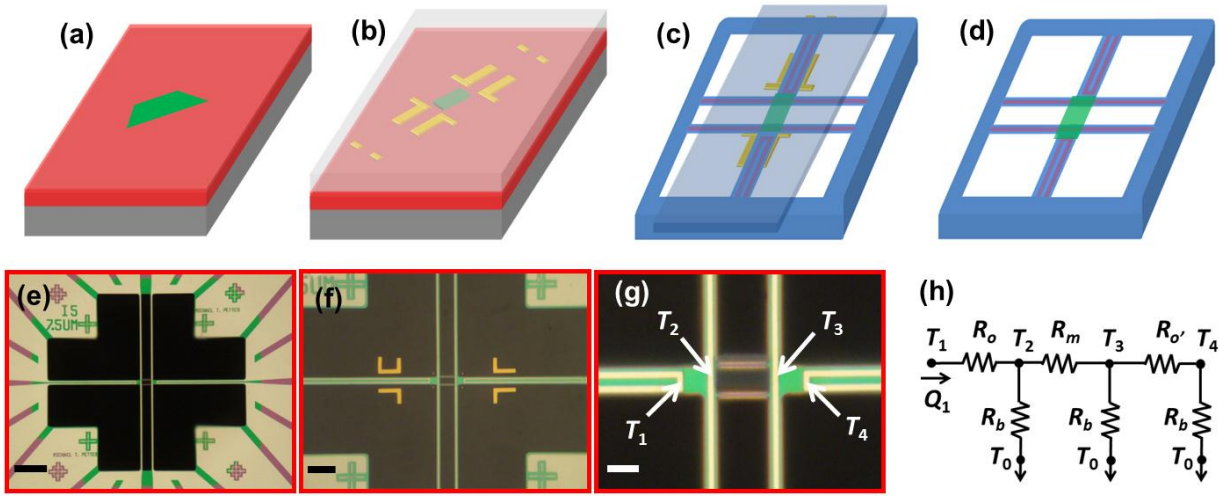

**Figure 1.** Transfer of exfoliated few-layer h-BN over a micro-bridge device. The top row consists of schematics showing (a) a few-layer h-BN flake (green) exfoliated on top of a Si substrate (gray) covered by thermal oxide (red), (b) Au registration marks (golden) on the substrate and a PMMA layer (semi-transparent) covering the patterned few-layer h-BN ribbon, (c) the PMMA carrier layer transferred on top of the micro-bridge device (blue), and (d) few-layer h-BN suspended on the central $SiN_x$ frame of the micro-device after the dissolution of the PMMA layer. The bottom row consists of optical micrographs showing (e) the micro-bridge device, (f) the PMMA layer aligned on the device, (g) an 11-layer h-BN sample suspended on the device after dissolving of the PMMA layer, and (h) equivalent thermal circuit of the measurement device, where $T_1$, $T_2$, $T_3$, and $T_4$ are the temperature at the middle point of the four Cr/Pt lines, as shown in (g). $R_m$ is the equivalent thermal resistance of the few-layer h-BN and the two $SiN_x$ bars underneath, $R_b$ is the thermal resistance of each of the four Cr/Pt/$SiN_x$ beams at the two ends of the h-BN sample, $R_o$ and $R_o$' are the thermal resistance of the two $SiN_x$ connections between the middle points of the U-shaped Cr/Pt line and the adjacent straight Cr/Pt line. The scale bars represent 25, 10, 5 μm in (e-g), respectively.

A through-substrate hole under the micro-bridge allowed for transmission electron microscopy (TEM) characterization of the crystal structure of the few-layer h-BN sample suspended on the device upon the completion of the thermal measurement. Table 1 lists the layer thickness and lateral dimensions obtained from TEM and scanning electron microscopy (SEM) analyses of each of the four h-BN samples measured in this work. The thickness of sample h-BN1 was determined from atomic force microscopy (AFM). For the other three samples, the layer thickness was obtained from the (0002) lattice fringes at the folded edge of each suspended h-BN sample. For example, the lattice fringes shown in Fig. 2(a) and 2(c) reveal that samples h-BN3 and h-BN4 consisted of 11 and 5 atomic layers, respectively. The selected area electron diffraction (SAED) patterns from the center of these samples produce only one set of Bragg reflections [Fig. 2(b,d)], indicative of a single crystal structure within the 4.25 μm diameter focused electron beam spot and AAA stacking order, which has also been observed in previous reports.[1, 13] Additionally, we consistently observed the presence of polymer residues on the surface of all four h-BN samples during TEM analysis, as shown in Fig. 2(a,c) and the supporting information. Similar polymer residue has been observed on the surface of suspended graphene or h-BN samples that were in contact with a PMMA layer during sample preparation.[14,

[15] Although thermal annealing under various gas environments has been employed to burn out the PMMA residue, it has been found that polymer residue cannot be completely removed from the sample surface.[15-19] In this work, annealing of the h-BN sample in air for 3 hours at 300°C was found to be ineffective in removing the PMMA residue.

**Table 1.** Dimensions of the suspended h-BN samples measured in this work.

|        | Number of Atomic Layers | Width ($\mu$m) | Suspended Length ($\mu$m) |
|--------|------------------------|----------------|---------------------------|
| h-BN1  | 12 ± 1                 | 9.0            | 3.0                       |
| h-BN2  | 12                     | 6.7            | 5.0                       |
| h-BN3  | 11                     | 6.5            | 7.5                       |
| h-BN4  | 5                      | 6.6            | 7.5                       |

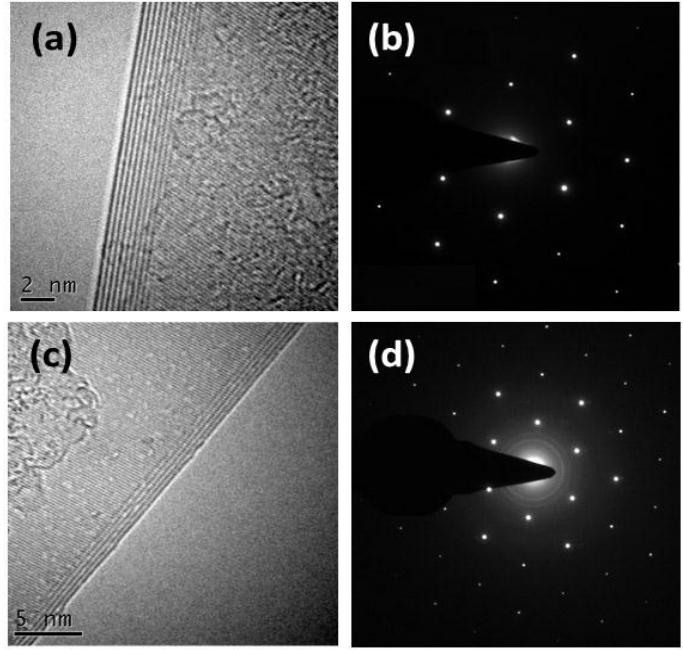

**Figure 2.** TEM images and electron diffraction patterns of few-layer h-BN samples. (a) TEM micrograph of the folded edge of sample h-BN3 showing that the sample consisted of 11 atomic layers. (b) SAED pattern taken from an 825 nm diameter region of sample h-BN3. (c) TEM micrograph of sample h-BN4 showing 5 atomic layers. (d) SAED pattern taken from a 4.25 μm diameter region of h-BN4.

The devices were placed in the evacuated sample space of a cryostat for thermal measurements. During the measurement, the electrical resistance of each of the four metal lines was measured to determine the average temperature rise ($\overline{\Delta T_j}$, $j$ = 1, 2, 3, and 4) in the metal line when one U-shaped Cr/Pt line was electrically heated. For the other U-shaped metal line and the two straight metal lines, the temperature rise at the mid-point of the metal line is twice of the average temperature rise, i.e. $\Delta T_j \equiv T_j - T_0 = 2\overline{\Delta T_j}; j = 2,3,4$, where $T_0$ is the substrate temperature. In addition, heat conduction analysis of the U-shaped heater line yields the temperature rise at its mid-point as $\Delta T_1 = \frac{3}{2}\overline{\Delta T_1} - \frac{1}{2}\left(\overline{\Delta T_2} + \overline{\Delta T_3} + \overline{\Delta T_4}\right)$.[12, 20] The thermal resistance values of the four Cr/Pt/SiN$_x$ beams ($R_b$) are designed to be identical, and can be found from the heat conduction analysis of the heater line in conjunction with the thermal resistance circuit in Fig. 1(h) as $R_b = 2(\Delta T_1 + \Delta T_2 + \Delta T_3 + \Delta T_4)/Q$, where $Q$ is the rate of electrical heating in the U-shaped heater line. Similar analysis yields the equivalent thermal resistance of the suspended h-BN and the two underlying SiN$_x$ bars as $R_m = R_b(\Delta T_2 - \Delta T_3)/(\Delta T_3 + \Delta T_4)$. The accuracy of this analytical solution has been verified by a three-dimensional finite element simulation.[20] The corresponding thermal conductance of the suspended h-BN and the two SiN$_x$ bars is then obtained as $G_m = 1/R_m$. The thermal conductance of the two SiN$_x$ bars ($G_n$) was measured using the same procedure when there was no h-BN on the device. The thermal conductance and thermal resistance of the suspended h-BN alone were then obtained as $G_s = G_m - G_n$ and $R_s = 1/G_s$, respectively. As shown in Fig. 3(a), the thermal conductance contribution of the two SiN$_x$ bars increased at low temperatures and was found to be in the range of 3-7% and 25-41% of the measured $G_s$ values for h-BN1 and h-BN4, respectively.

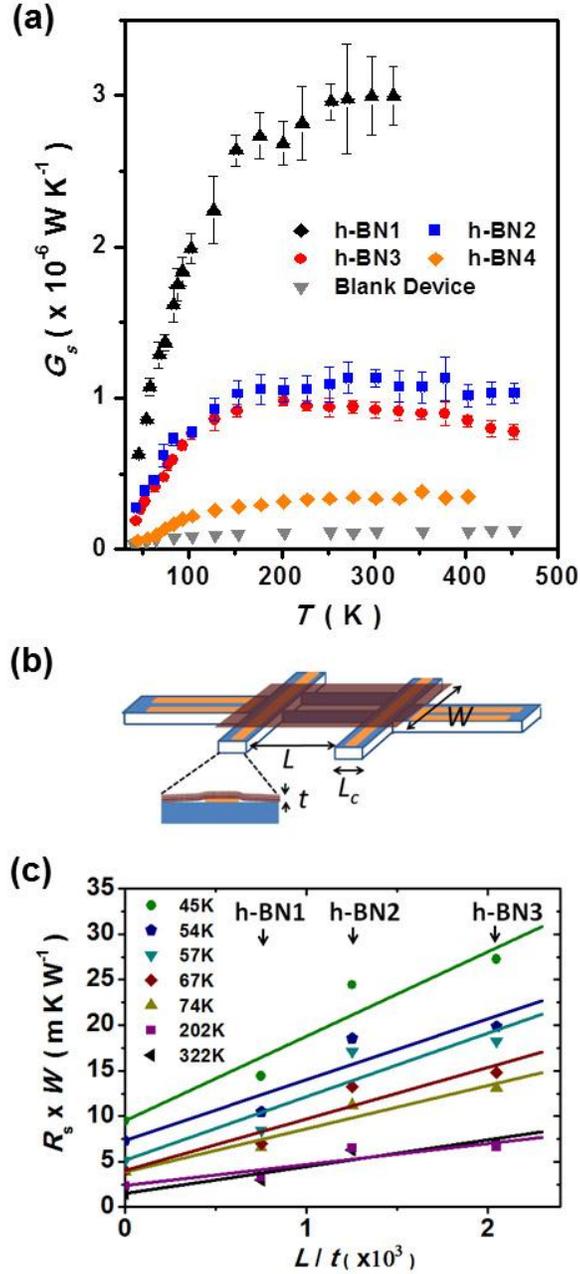

**Figure 3.** (a) Measured thermal conductance values of four h-BN samples ($G_s$) and of the two 7.5 μm long $SiN_x$ bars ($G_n$) of one blank device. (b) Schematic showing the relevant dimensions of an h-BN sample suspended on the central frame of the micro-bridge device. (c) The $R_sW$ product versus the $L/t$ ratio for the three 11-12 layer thick samples at different temperatures indicated in the legend. The lines are linear fitting to the measurement values shown as symbols.

The obtained thermal resistance of the h-BN sample contains the diffusive thermal resistance ($R_d$) and the contact thermal resistance ($R_c$), i.e., $R_s = R_d + R_c$ and $R_d = L/(\kappa t W)$, where $\kappa$, $L$, $t$, and $W$ are the thermal conductivity, length, thickness, and width of the suspended h-BN, as illustrated in Fig. 3(b). The $R_d$ value decreases with increasing $t$ and decreasing $L$ so that $R_c$ can be appreciable for thick h-BN samples with a short suspended length. In this work, $L$ varies from 3 to 7.5 μm for the three 11-12 layer thick samples. Based on the assumption of the same contact resistance per unit contact width for all samples that have a similar contact length ($L_c$) as well as the assumption of the same thermal conductivity for the three samples with a similar layer thickness, we use the relation, $R_s W = L/(\kappa t) + R_c W$, to fit the $R_s W$ versus $L/t$ data of the three 11-12 layer samples, as shown in Fig. 3(c). The as-obtained $R_c W$ product, which corresponds to the intercept of the linear fitting at $L = 0$, is used to determine the $R_c$ value for the two 7.5 μm long samples with a thickness of 11 and 5 layers, respectively. The obtained $R_c$ value increases from 20% to 33% of the measured $R_s$ for the 11-layer sample when the temperature decreases from 300 K to 45 K. In comparison, the obtained $R_c$ is only 8% and 11% of the measured $R_s$ for the 5-layer sample at temperature 300 K and 45 K. In both cases, the obtained $R_c$ is subtracted from the measured $R_s$ to obtain the $R_d$ and $\kappa$ values of the two samples.

As shown in Fig. 3(c), the measurement data scatter around the linear fitting line, which can result in uncertainties in the as-obtained $R_c$ and $\kappa$. One possibility for such scattering is caused by the variation of $R_c W$ for three different samples because of the different contact lengths between the h-BN and the 3.8 μm-wide beam with the U-shape metal line. To address this variation of the contact length, we have developed another contact thermal resistance model, where the interface thermal conductance per unit area ($g_i$) instead of the $R_c W$ product is assumed to be constant. As discussed in detail in the Supporting Information, this second model yields similar results as those obtained by the simple linear fitting model. Both models suggest that $g_i$ increases from about 1.7-2.7 ×10$^5$ Wm$^{-2}$K$^{-1}$ at 45 K to about 1.7-2.4 ×10$^6$ Wm$^{-2}$K$^{-1}$ at 300 K (see Supporting Information). In general, $g_i$ is expected to increase with increasing temperature because it is proportional to the specific heat that increases with temperature. However, the $g_i$ values from our measurements are 1-2 orders of magnitude lower than those reported for graphene-SiO$_2$[21] and graphene-metal interfaces,[21, 22] likely because of roughness on the micro-bridge device and contamination left at the interface between the as-transferred few-layer h-BN and the device.

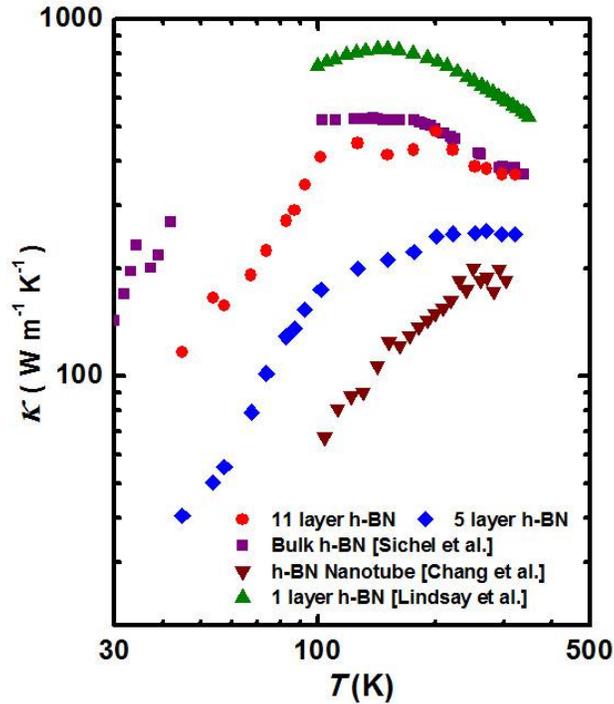

**Figure 4.** Thermal conductivity values of the two 7.5 μm long, 11 and 5 layer thick suspended h-BN samples as a function of temperature. Shown in comparison are the measured basal plane thermal conductivity reported by Sichel et al. for bulk h-BN[4] and by Chang et al. for a multi-

walled h-BN nanotube,[22] and the calculation result reported by Lindsay et al. for a single-layer suspended h-BN.[9]

As shown in Fig. 4, the obtained room-temperature thermal conductivity of the 11-layer sample is about 360 Wm$^{-1}$K$^{-1}$, which approaches the basal-plane value reported for bulk h-BN.[4] As the temperature decreases, the thermal conductivity of this sample initially increases due to reduced Umklapp scattering, reaching its peak value between 100 K and 200 K, similar to bulk samples. At lower temperatures, the thermal conductivity of the 11-layer sample becomes lower than the bulk values. A recent theoretical calculation has suggested that the thermal conductivity of atomically-clean suspended few-layer h-BN increases as the number of layers decreases because of the reduction of interlayer phonon scattering.[9] Our results do not follow this trend as the thermal conductivity of the 5-layer sample is lower than that of the 11-layer sample. In addition, the thermal conductivity suppression in the thinner sample becomes more pronounced at lower temperatures with the peak value appearing at near room temperature. Interestingly, both samples measured here maintain higher thermal conductivity than that reported for a multi-walled h-BN nanotube, which was measured with a different micro-bridge device without eliminating the contact thermal resistance.[23]

To understand the thermal conductivity suppression at low temperatures, we first evaluate the impact of phonon scattering by the lateral edges of the h-BN ribbon and by point defects. We have calculated the thermal conductivity of the suspended h-BN according to a solution of the phonon Boltzmann transport equation,[19]

$$\kappa = \frac{1}{4\delta} \sum_p \int_{\omega=0}^{\omega_{max}} v_p^2 \tau_p D_p(\omega) \hbar\omega \frac{d<n>}{dT} d\omega \qquad (1)$$

where $\delta$ is the interlayer spacing, $p$ is the phonon polarization, $v_p$ is group velocity, $\tau_p$ is the relaxation time, $D_p(\omega)$ is the density of states per unit area, and $<n>$ is the Bose-Einstein distribution. The phonon dispersion of bulk h-BN was obtained from an *ab initio* calculation based on local density approximation with the use of Quantum ESPRESSO. The phonon scattering rate by the lateral edge of the few-layer h-BN samples was calculated as $\tau_{B,p}^{-1}=v_p/l_b$, where $l_b$ is the edge scattering mean free path. The boron atoms in our samples consist of 19.9% $^{10}$B and 80.1% $^{11}$B. Phonon scattering by the isotopic impurities was considered in the calculation. Due to the high concentration of isotope impurities, other atomic impurities or structural defects such as point vacancies have been found to play a less important role than isotope impurity scattering even when the point vacancy concentration is as high as $10^{20}$ cm$^{-3}$. Without accounting for scattering by polymer residues and Umklapp scattering, the latter of which is negligible at low temperatures, the calculated thermal conductivity shows good agreement with the measurement results at temperatures below 100 K when $l_b$ is taken to be 550 nm and 180 nm for the 11 and 5 layer h-BN samples, respectively (See Supporting Information). However, the values are much smaller than the 3–7.5 μm length of the single-crystalline h-BN samples and the 3.5 μm separation between the two SiN$_x$ bars below the h-BN. Hence, edge scattering and point defect scattering are not responsible for the considerably suppressed thermal conductivity at low temperatures.

Several theoretical calculations have suggested that the presence of surface functional groups can result in large suppression of the thermal conductivity of suspended graphene.[24, 25] Similarly, the presence of the polymer residue on the sample surfaces can lead to even stronger suppression. Such scenario has been suggested in our recent measurement of suspended bi-layer graphene.[19] Because the polymer residue was clearly observed from the TEM images on the surfaces of the h-BN samples as a result of fabrication process, we attribute the trend of decreasing thermal conductivity with decreasing layer thickness to phonon scattering by the polymer residue. Such scattering is expected to play an increasing role as the layer thickness decreases. In particular, because low-frequency phonons dominate the low-temperature thermal conductivity, the larger suppression found in the thinner sample at lower temperatures suggests a higher scattering rate of low-frequency phonons by the polymer residue. At high temperatures, on the other hand, polymer-induced scattering becomes less important compared to Umklapp scattering, thus the suppression from bulk thermal conductivity becomes smaller, especially in the thicker samples.

Our measurement results show that the room-temperature thermal conductivity of suspended h-BN can approach the basal-plane values of bulk h-BN crystals at room temperature when the thickness increases to more than 10 atomic layers in spite of the presence of polymer residue on the sample surface. As the sample thickness decreases, the thermal conductivity decreases because of increasing phonon scattering by polymer residues, which is more pronounced at low temperatures or for low-frequency phonons. These new findings, especially the thickness needed to obtain the bulk thermal conductivity values in these few-layer h-BN samples with polymer residues, provide new insight into the effect of polymer residues and surface functional groups on phonons and especially low-frequency phonons in two-dimensional layer materials. The result is of value for the exploration of few-layer h-BN as a heat spreading layer in novel electronic devices made of flexible polymeric substrate, or as fillers to enhance the thermal conductivity of polymeric composites.

**Corresponding Author**

* E-mail: lishi@mail.utexas.edu

ACKNOWLEDGMENTS

The authors acknowledge financial support for related research from the United States Department of Energy, award number DE-FG02-07ER46377, and from the Office of Naval Research, award number N00014-10-1-0581.